\documentclass[12pt]{article}
\usepackage{amsmath}
\usepackage{graphicx}
\usepackage{enumerate}
\usepackage{natbib}
\usepackage{color}
\usepackage{soul}
\usepackage{url} % not crucial - just used below for the URL
\usepackage{ amssymb, graphicx, multirow,amsthm,gensymb}
  \theoremstyle{definition}
  \newtheorem{defn}{\protect\definitionname}
 \theoremstyle{definition}
\usepackage{babel}
  \providecommand{\definitionname}{Definition}

\newcommand{\blind}{1}

\begin{document}

\def\spacingset#1{\renewcommand{\baselinestretch}%
{#1}\small\normalsize} \spacingset{1}

%%%%%%%%%%%%%%%%%%%%%%%%%%%%%%%%%%%%%%%%%%%%%%%%%%%%%%%%%%%%%%%%%%%%%%%%%%%%%%

\if1\blind
{
  \title{\bf Differential Privacy for Government Agencies---Are We There Yet?}
  \author{J\"{o}rg Drechsler\\%\footnote{J\"{o}rg
    %Drechsler is distinguished researcher, Institute for Employment Research,
    %Department for Statistical Methods, Regensburger Stra{\ss}e
    %104, 90478 N\"{u}rnberg, Germany, and Associate Research Professor, Joint Program in Survey Methodology, University of Maryland, College Park
%1218 Lefrak Hall College Park, MD. 20742(e-mail: joerg.drechsler@iab.de)}\\
    Institute for Employment Research and University of Maryland}
    \date{}
  \maketitle
} \fi

\if0\blind
{
  %\bigskip
  \bigskip
  \bigskip
  \begin{center}
    {\LARGE\bf Differential Privacy for Government Agencies -- Are We There Yet?}
\end{center}
  \medskip
} \fi

\bigskip
\begin{abstract}
Government agencies typically need to take potential risks of disclosure into account whenever they publish statistics based on their data or give external researchers access to collected data. %For this reason, research on disclosure avoiding techniques has a long tradition at statistical agencies.
In this context, the promise of formal privacy guarantees offered by concepts such as differential privacy seems to be the panacea enabling the agencies to quantify and control the privacy loss incurred by any data release exactly. Nevertheless, despite the excitement in academia and industry, most agencies---with the prominent exception of the U.S. Census Bureau---have been reluctant to even consider the concept for their data release strategy.

This paper discusses potential reasons for this. We argue that the requirements for implementing differential privacy approaches at government agencies are often fundamentally different from the requirements in industry. This raises many challenges and questions that still need to be addressed before the concept can be used as an overarching principle when sharing data with the public. The paper does not offer any solutions to these challenges. Instead, we hope to stimulate some collaborative research efforts, as we believe that many of the problems can only be addressed by interdisciplinary collaborations.
\end{abstract}

\noindent%
\vfill

\newpage
\spacingset{1.5} % DON'T change the spacing!
\section{Introduction}
The concept of differential privacy (DP) has gained substantial attention in recent years as the only standing approach offering formal privacy guarantees that hold irrespective of the assumptions regarding the background knowledge of a potential attacker. Since the seminal paper by \citeauthor{Dwork2006} was published in 2006, thousands of papers---mostly from the computer science community---have addressed the topic from various perspectives. Given its roots in theoretical computer science, it is perhaps not surprising that most of these works approach the problem from a theoretical perspective. While new algorithms that satisfy the DP requirements for specific analysis tasks are proposed almost daily, their impacts on accuracy are typically only evaluated analytically by looking at measures such as the maximum expected error under asymptotic regimes% (that is, assuming $n\rightarrow\infty$)
. These metrics only offer limited insights for practical applications on real data with a fixed sample size. %Besides, evaluations based on small sample sizes are sparse, although it is well understood that the relative advantages of different algorithms crucially depend on the size and structure of the available data (see \cite{alabi2020} for an illustration in the context of linear regression).
Furthermore, problems that commonly arise in practice, such as complex data structures, missing data, data cleaning, and so on tend to be ignored.

Although practical experience has been relatively limited so far, the concept of differential privacy has been embraced by (some parts of) industry in recent years. Several companies, especially from the tech industry, such as Google \citep{erlingsson2014}, Apple \citep{adp2017}, Microsoft \citep{ding2017}, Facebook \citep{Facebook2020}, and Uber \citep{Uber2017}, have deployed the concept for some of their products or are currently conducting research with the aim of adopting the approach in the future.

Despite the excitement in academia and industry, the enthusiasm at government agencies and national statistical organizations (NSOs) has so far been limited. While some agencies explored the feasibility of the approach in limited settings \citep{Soria2013,bailie2019}, the only large-scale deployment of the approach for many years was OnTheMap, a graphical user interface offered by the U.S. Census Bureau visualizing commuting patterns in the United States. The underlying data are protected using an algorithm satisfying ($\epsilon$,$\delta$)-probabilistic DP \citep{machanavajjhala2008}. More recently, the concept has also been adopted by the U.S. Census Bureau to release earnings percentiles of graduates from post-secondary institutions \citep{Foote2019}.

Possibly the most prominent deployment of DP at government agencies so far was the adoption of the concept for the 2020 Decennial Census, which was first announced in 2018 \citep{abowd2018}. Compared to most other data products gathered at NSOs, which are based on surveys with limited sample sizes and hundreds of variables, protecting data from the decennial census seems to be a straightforward task: it contains more than three hundred million records at the individual level \citep{Census2021} and only asks up to seven questions at the individual level and three questions at the household level \citep{Census_quest}. Nevertheless, the fact that it took a research team of computer scientists and statisticians several years to develop a suitable DP algorithm and the concerns regarding the accuracy of the results that were raised after results from a test run of the algorithm using 2010 Census data were released \citep{ruggles2019,Mervis2019,van2020,santos2020,Maine_letter2020,winkler2021} illustrate the difficulties involved in trying to implement the ideas in practice.

Of course, these concerns need to be put into perspective. On the one hand, the different test runs released by the U.S. Census Bureau in 2020 had a strong focus on ensuring a high level of privacy. The privacy parameters used for publishing results from the decennial census are much higher than those used for the test run \citep{Census_parms}, which implies that these results will be substantially more accurate than those from the test runs, but also offer less protection (see Section \ref{sec:dp_intro} for a discussion of the privacy parameters and their interpretation). On the other hand, some of the concerns only arise as some users of the data realize for the first time that the numbers published based on the decennial census are not free from error \citep{garfinkel2018}. Other sources of error stemming from nonresponse, undercoverage, imputation, and previous strategies used to protect the data, such as swapping, have largely been ignored in previous analyses \citep{Hawes2020}. Arguably, some of the very detailed analyses that were conducted based on earlier censuses never gave reliable results, but since quantifying the magnitude of these other sources of error is difficult, the uncertainty and potential bias in the results were never taken into account (see \citet{boyd2019} for a more detailed discussion on this point).

One of the key challenges which distinguish the decennial Census from previous deployments in the tech industry is that (accurate) answers are required on a very detailed geographical level and for billions of statistics, which implies that despite its large size, many of the statistics released are based on a very limited number of cases and an enormous amount of very detailed output needs to be protected. Many additional problems arise because the requirements when implementing DP at government agencies are fundamentally different from the requirements in industry. Some of the key challenges that will be discussed in this paper include: (i) usage of the data for multiple purposes by multiple analysts over extended periods of time, often based on limited sample sizes (Section 3); (ii) difficulties in anticipating the impacts on accuracy for unforeseen analyses and challenges in fully understanding the privacy guarantees provided (Section 4); (iii) analysis goals that are different and more complex than in the industry context (Section 5); (iv) methodological challenges when assessing the impacts of complex data collection routines and various pre-processing steps such as weighting, editing, and imputation on the privacy guarantees (Section 5); (v) challenges in finding the right value for $\varepsilon$ (Section 6).
%The amount of data is much more limited, the data should be available for many years, results should be reproducible, users of the data are typically interested in making inferences regarding a specific target population, agencies are not the final users of the data, incentives for sharing the data are virtually non-existent, etc.
All of these aspects need to be taken into account when considering whether or not the concept might be a viable approach for solving the ever-present dilemma between confidentiality protection and broad access to the data.

This paper is not intended to provide a road map for how to implement DP at government agencies. Nor will it discuss the advantages and disadvantages of adopting formal privacy methods relative to the more traditional methods commonly employed at statistical agencies. In this regard, it is also important to emphasize that several of the challenges discussed in this paper also apply to many of the traditional methods. %Applications of differential privacy at government agencies are only beginning to emerge in recent years and many challenges still need to be solved, especially if this concept should also be adopted in the context of survey data.
One of its major goals is to %make readers aware of these challenges to stimulate research that
look at the concept from a statistical and applied perspective. Most of the research conducted so far has occurred in the field of theoretical computer science. Substantial progress has been made in developing the theoretical underpinnings of the concept, but statistical inference and practical problems with real applications only play a minor role in most of the previous research. This needs to change if the concepts of differential privacy are to be adopted more widely by government agencies. %However, computer scientists cannot be expected to solve these challenges alone as these problems are too far from their own research.
Thus, the article aims at summarizing the challenges that remain with the hope that statisticians and researchers from other fields will step up and collaborate in working on these problems. %The aim of the paper is to highlight some important aspects that need to be considered and open questions that still need to be addressed when thinking about if and how the concept of differential privacy could be applied in the government context.

The remainder of the paper is structured as follows: Before discussing the different aspects that need to be considered in the government context, Section \ref{sec:dp_intro} provides a short review of DP and some of its properties. Section \ref{sec:data} discusses the challenges arising concerning the typical data products at government agencies with limited sample sizes but very detailed information. It also illustrates why establishing a query response system or %, which is a popular setting for the deployment of differential privacy, nor
providing restricted access to the unprotected data for accredited researchers seems to be problematic when adopting DP for government agencies. Section \ref{sec:priv_accuracy} addresses the difficulties in anticipating the impacts of DP on the accuracy of the results obtained and highlights the fact that understanding the level of protection provided through DP is not a trivial matter. Differential privacy in the survey context is the focus of Section \ref{sec:surveys}, which illustrates potential negative effects on response rates and discusses challenges when trying to incorporate common data processing steps such as weighting and imputation. It also discusses problems when trying to account for the data protection procedures when making inference to the underlying population. Section \ref{sec:setting_eps} addresses the difficulties that government agencies face when setting the privacy parameter $\varepsilon$. The paper concludes with some suggestions for addressing the challenges and open questions raised, advocating for more interdisciplinary research to tackle these problems.

\section{A brief review of differential privacy}\label{sec:dp_intro}
This section will only review the definition and properties of $\varepsilon$-DP as originally proposed by \citet{Dwork2006}. Meanwhile, various relaxations of the initial concept---most notably ($\varepsilon$,$\delta$)-DP---have been proposed in the literature. Since the subtleties of these variants are irrelevant for the discussions in the remaining paper, we refer interested readers to \citet{dwork2014} and \citet{vadhan2017} for a more detailed introduction to DP and its variants.%, which also cover (some of the) relaxations of $\varepsilon$-differential privacy.

A popular context taken to illustrate the underpinnings of DP is a query response system. A system of this kind accepts specific queries as input---a query for the mean of a variable, for example---and then returns a noisy answer to the query, with the noise calibrated to ensure that the requirements of DP are met. %The user of the system never accesses the underlying data directly, she will only see the noisy answer to the query.
In this setting, differential privacy guarantees that the influence that any record in a database can have on the reported output is strictly limited. This ensures that the information that can be learned about any individual in the database is also limited. These guarantees are formalized by bounding the difference of the probability distribution
of the query response when changing one record in the data.

\begin{defn}[$\varepsilon$-differential privacy, \cite{Dwork2006}]
\label{differential_privacy}A randomized mechanism $\mathcal{M}$ gives
$\varepsilon$-differential privacy if, for all neighboring datasets $D$,
$D'$, and all events $S\subset Range(\mathcal{M})$
\begin{equation}
P(\mathcal{M}(D)\in S)\le e^{\varepsilon}P(\mathcal{M}(D')\in S)\label{eq:diff_priv}.
\end{equation}
\end{defn}

Depending on the context, two datasets are called neighboring, if one could be obtained from the other by (a) adding or removing a single record (unbounded DP) or (b) changing the values of one record, while keeping the size of the database fixed (bounded DP).

Differential privacy ensures that the probability of observing a specific output if database $D$ is used as the input is never more than $e^{\varepsilon}$ times (and never less than $e^{-\varepsilon}$ times) the probability of observing the same output if database $D'$ is used as the input, where the probability distribution is based on the randomness induced by the mechanism $\mathcal{M}$. The parameter $\varepsilon$ can be used to specify the level of protection. Larger values of $\varepsilon$ allow for larger differences in the output distribution between two neighboring databases, thus offering lower levels of protection. However, larger values of $\varepsilon$ will typically increase the level of accuracy of the reported output as the mechanism $\mathcal{M}$ will need to introduce less noise to satisfy Equation (\ref{eq:diff_priv}). Thus, $\varepsilon$ can be seen as a tuning parameter that trades privacy for accuracy of the estimate obtained.

The concept of differential privacy is attractive for several reasons. First, it offers strong (and mathematically proovable) privacy guarantees as it quantifies the risks from joining the database in a worst-case scenario, considering all possible databases differing in one record. These strong guarantees are especially important, as recent research has demonstrated that previously employed strategies such as cell suppression might not protect the data sufficiently \citep{Census2021reconstruction,garfinkel2019}. Second, the underlying idea is very intuitive: if the probabilities of obtaining a specific result will only change marginally, whether my data are in the database or not, I can feel confident that little can be inferred about me based on this result. Third, the guarantees hold irrespective of any assumptions regarding the knowledge of a potential attacker. This is an important advantage over risk assessment strategies that are currently used at statistical agencies. These strategies typically start by making assumptions regarding information that an attacker possesses about the units contained in the database and then try to assess the disclosure risks based on this assumption. Clearly, these risk assessments will fail if the assumptions are not correct, more information becomes available in the future, or the attacker uses a different attacking strategy than those anticipated by the agency (see \citet{reiter2019} for further discussion of this point). Fourth, unlike data access regulations currently being implemented, DP makes the ever-present trade-off between accuracy and privacy explicit \citep{Hawes2020}. Any data protection strategy, from aggregating geographic information to top-coding to swapping sensitive values to not releasing the data at all, makes implicit decisions on trading accuracy against privacy. However, the exact trade-off is more difficult to understand and control under current regimes. With DP it is conceivable (at least in theory) to select the level of accuracy and privacy that is optimal for society. Finally, since the protection strategy is fully transparent, that is, the selected value of $\varepsilon$ and any information about the algorithm used to protect the data can be publicly released, it would be possible in principle to take the uncertainty induced by the algorithm into account when analyzing the protected data. This is difficult with protection methods currently being employed, such as swapping or noise addition, since their privacy parameters, that is, the swapping rate or the variance of the noise term, are typically treated as confidential \citep{reiter2019}. %Still, finding the optimal value for $\varepsilon$ and integrating the extra uncertainty in downstream analyses are daunting tasks in practice (see also the discussions in Sections \ref{sec:setting_eps} and \ref{sec:valid_inference}).
%Finally, enforcing differential privacy might have the positive side effect of analysts thinking more carefully about the robustness of their findings. Since the amount of noise that needs to be infused to ensure differential privacy depends on the sensitivity of the statistic of interest to small changes in the database, analysts will likely adopt robust estimation strategies whenever possible. This might also provide some robustness against all of the other sources of error in the data that are typically ignored \citep{boyd2019}.

Differential privacy offers two additional attractive properties:% that will be relevant for the discussions in the remainder of this paper:
\begin{itemize}
\itemsep-5pt
\item
\textbf{Postprocessing:} Differential privacy is immune to postprocessing, meaning that any function of an output satisfying $\varepsilon$-DP also satisfies DP with the same level of $\varepsilon$.
\item
\textbf{Composition:} If the same database is used to answer $K$ queries in a DP manner, each with its own privacy parameter $\varepsilon_k$, $k=1,\ldots,K$, the overall privacy loss is bounded by $\varepsilon=\sum_{k=1}^K \varepsilon_k$.
\end{itemize}

%These properties have several important implications in the government context: The first property guarantees that both the data producer as well as the user of the data can arbitrarily alter the output without increasing the privacy loss. This implies that the data disseminating agency can adjust the differentially private output to avoid releasing implausible values such as negative age values and to improve the utility. The agency also does not have to worry that a malicious user of the data might be able to learn sensitive information by manipulating the received information in a clever way. This property holds even if the output is combined with information from other data sources.

%The second property allows one to split the overall privacy budget $\varepsilon$ across multiple queries. In principle this would allow one to assign individual privacy budgets to different users of the data. The users could then decide how to best spend their budget to obtain the information they are interested in (but see the caveats discussed in Section \ref{sec:no_query_response} below).

We refer the interested reader to \citet{dwork2014} for an in-depth discussion of the properties of DP including several examples of available mechanisms that meet the requirements of DP for various analysis tasks.

\section{Data availability and access}\label{sec:data}
Many of the previous deployments of DP used the concept in situations where massive amounts of data are collected every day \citep{erlingsson2014,adp2017,ding2017,Facebook2020}. This offers two advantages: First, for certain types of queries the sensitivity of the statistic of interest, which governs how much uncertainty needs to be introduced to ensure DP, does not depend on the size of the data (sensitivity in this context measures how much the statistic changes if one record in the data is altered). One example of such queries are counting queries, that is, statistics that simply count how many records in a database have a specific set of attributes. %, for example the number of users of a specific internet browser at a specific time of the day (see, for example, Section 3.3 in \citet{dwork2014} for further discussion of counting queries and their properties in the DP context).
This implies that the same amount of noise needs to be added to protect a frequency table (which is based on multiple counting queries) irrespective of whether the table is based on 20 records or 10 million records.
%However, adding, say, a noise value of 3 to a cell count of 5 will obviously have a stronger impact on any findings based on this table than adding the same value to a cell count of 3.5 million. In other words, the signal clearly dominates the noise in the latter case, but not in the first case.
Second, %the data can become irrelevant quickly with regards to the purpose for which they are collected. With so much data being collected everyday, there is often no need to still look at yesterday's data. Furthermore,
the companies typically use the data to learn about current user behavior %to optimize their products or to generate revenue from targeted advertising based on the collected information. For these purposes, long term trends in user behaviour are often irrelevant
and using older data will increase the risks of not adopting fast enough to changes in this behavior. This implies (at least in theory) that one can start with a new privacy budget regularly since older data quickly lose their value (there is a caveat, as more data about the same individuals might be collected over time, which will spill additional information, but this issue seems to be ignored, at least in some of the deployments \citep{tang2017,josep2020}).
%The sensitivity basically measures, how much the statistic of interest can change, if we change one record in the data. To illustrate this point, we can look at frequency tables, which have a low level of sensitivity (the sensitivity is one or two, depending on whether we consider changing one record to mean adding or removing on record (unbounded differential privacy) or changing the values of one record (bounded differential privacy)).

\subsection{Data collected at government agencies}
While large-scale administrative data and other massive data sources such as satellite images or cell phone data have been exploited more extensively in recent years, most of the information collected at government agencies is still based on surveys. Most of these surveys comprise fewer than 100,000 records and often collect very detailed information on every unit included in the database. Thus, compared to previous deployments of DP, databases at government agencies will typically contain fewer cases but more information, making the release of accurate privacy-protected results more challenging \citep{reiter2019}.

 At the same time, the data will remain relevant over many years. %(given the substantial costs of data collection any thing else would be unacceptable).
The fact that the U.S. Census Bureau is required by law to make census data publicly available 72 years after the initial data collection indicates that there is still interest in these data even more than 70 years after they were collected. In fact, \citet{IPUMS2021} lists more than 150 articles and books that used data from the 1940 Decennial Census since its public release in 2012. However, this implies that the privacy budget, that is, the level of $\varepsilon$ spent for all information that is released, needs to protect the data for all of these years.

\subsection{Query response system not an option}\label{sec:no_query_response}
As pointed out above, the concept of differential privacy is often discussed in the context of a query response system. Such systems are prevalent at statistical agencies and typically referred to using terms such as \textit{remote analysis servers} (RAS). With RAS users will never see the actual microdata. Instead, they will define the analysis of interest by selecting the type of analysis to be performed and the variables to be included from various drop-down menus. The RAS will compute the analysis based on the underlying microdata but will only return the (potentially perturbed) results. Examples of this include Real Time Remote Access  (\url{https://www.statcan.gc.ca/eng/rtra/rtra}) provided by Statistics Canada, which allows user-defined tabulations, and the Cross-National Data Center in Luxembourg (\url{www.lisdatacenter.org}), which additionally allows for simple regression models to be specified. We note that none of the current implementations of RAS offer formal privacy guarantees at this point. Instead, some of them rely on ad hoc suppression rules and may be vulnerable to disclosure attacks such as those discussed in \citet{Census2021reconstruction} and \citet{garfinkel2019}.

In theory, dealing with multiple queries that are answered by an RAS over time is straightforward in the context of DP due to the composition property described above. If the overall privacy loss deemed to be acceptable is $\varepsilon$ and it is known that $k$ queries should be answered, one option would be to assign $\varepsilon/k$ of the privacy budget to each query.
However, in practice, a dynamic query system in which many queries are submitted by different users raises many challenges. Distributing the privacy budget equally across queries, for example, is not necessarily the optimal solution or even the fairest \citep{reiter2019}. Two queries spending the same level of $\varepsilon$ might have completely different signal-to-noise ratios depending on the type of query and the sample size used for analysis. To achieve the same level of $\varepsilon$, for example, much more noise typically needs to be added when computing the mean compared to computing frequency counts. As pointed out by \citet{garfinkel2018} ``(d)ifferential privacy lacks a well-developed theory for measuring the relative impact of added noise on the utility of different data products.'' Furthermore, spending the same privacy budget on some intermediate analysis during data preparation as on estimating measures of political significance such as the poverty rate, which will impact the allocation of billions of dollars, would appear questionable. Thus, deciding how to split the privacy budget among multiple queries is a daunting task \citep{reiter2019}. %Note that the U.S. Census Bureau faced a similar dilemma, when deciding how to split the privacy budget across the myriad of tables that are published based on the Decennial Census \citep{Hawes2020}.

Nevertheless, these discussions are based on the assumption that all queries are known in advance. The main motivation for providing broad data access for the scientific community and the general public, however, is the understanding that statistical agencies cannot anticipate all potential research questions in regard to which the database might provide useful insights. If this were the case, the agency could run all of these analyses once the data had been collected and publish all of the results (potentially under the constraints of DP). Based on this assumption, no external access would be required. However, once we accept that users will submit queries that have not been anticipated, the difficult question to answer is: how much of the privacy budget should the agency hold back to be able to answer these questions? If the agency is too restrictive, answers to the submitted queries will be unnecessarily inaccurate, but if the agency is too generous, there will be a point at which all the privacy budget has been spent. At this point, the system has to be shut down forever and no one can ever be allowed to access the data again unless a decision is made to lower the privacy standards at some point in time based on the argument that older data need less protection (the public release of U.S. Census data after 72 years is one example of such a decision). Alternatively, it could be argued that previously unanticipated analyses are so beneficial to society that additional sacrifices of privacy seem justified.

It is certainly true that from a theoretical perspective the amount of information that can be gleaned from any data is limited, and once a sufficiently large number of statistical analyses have been conducted, any additional analyses are only a (potentially complex) function of the previous results. At some point, therefore direct access to the data would no longer be required. The famous reconstruction theorem of \citet{dinur2003} is a natural extension of this logic: there is only a limited number of statistical analyses that can be published unaltered before the entire database can be reconstructed from the published results. Still, using this reasoning to argue that access to the data is no longer necessary if enough previous analyses are available is based on two strong assumptions: 1) potential users of the data are aware of all these previous analyses and 2) they are able and willing to derive how to use all of these previous findings to an optimum degree to answer the question that is of interest to them. These assumptions are unlikely to hold in practice. Instead, this policy would likely set up incentives to submit queries as early as possible. This would increase the risk of sloppy analysis, as researchers might be tempted to avoid spending time and the privacy budget on robustness checks and careful evaluations of the modeling assumptions. Furthermore, important research questions might only emerge at a time when all of the budget had already been spent. A side from this, replication studies to validate earlier findings would be challenging, especially as carefully designed replications require the results to be evaluated based on various assumptions to check the robustness of the findings. This could imply that these studies might require a larger privacy budget than the initial research. Finally, given that most of the data hoisted by statistical agencies are collected using taxpayers' money, it would be difficult to sell the idea that no one would ever be allowed to access the data again after the privacy budget had been spent.

Given all of these observations, it seems that the only sensible strategy for statistical agencies willing to adopt the approach will be to generate a differentially private copy of the microdata and disseminate these microdata to the public. Given that DP is immune to postprocessing, the users would be able to run as many analyses on these data as they wanted without violating the DP conditions. The only requirement would be that the original data would never be touched again, as any additional information released based on the original data would require a new privacy budget (unless some of the privacy budget had been withheld---for validation purposes, for example). %We note that generating differentially private microdata is also the strategy followed by the U.S. Census Bureau for the decennial Census.
Nevertheless, intuitively speaking, it is more difficult to protect the entire microdata (input perturbation) than only specific statistics (output perturbation). A similar point is made in \citet{josep2020}, who discuss the fundamental challenges involved when using DP for microdata release. %Thus, the answers obtained from the differentially private microdata will necessarily be noisier than the answers that could be obtained if an optimal differentially private algorithm could be used to answer all the queries of interest.
With the high dimensional survey data discussed in the introduction in particular, the answers obtained from the protected data might be so noisy that they would no longer provide any useful insights. It is conceivable that algorithms for generating microdata could be developed that provide very accurate answers for specific types of queries, but in order to achieve this, these queries would need to be known in advance, which would leaf to the same problems already discussed in regard to the query response system.

\subsection{Problems with tiered access for accredited researchers}
One potential solution to circumvent the difficulties of differentially private microdata would be to allow trusted researchers access to the unprotected microdata on the premises of the statistical agency. They would only need to ensure that any result that they published satisfied the requirements of DP. This approach is in line with current data access regulations at many agencies. To facilitate access, many NSOs established research data centers (RDCs). Compared to the microdata samples that are disseminated to the public, there are datasets available at the RDCs that are more detailed and less protected. %Access is only granted to accredited researchers, the researchers cannot bring any own devices to the RDC, are often monitored while working at the RDC, and any research output can only be used outside the RDC after the output has been carefully scrutinized by employees of the RDC to ensure that it does not violate any confidentiality constraints.

Maintaining tiered access of this kind in the context of DP would offer two important advantages: First, fitting the final model of interest typically forms only a very small part of an applied research project. Most of the work goes into data preparation and data cleaning, checking model assumptions, and so on. All of these steps could be performed without spending privacy budget, leaving more of the budget to obtain more accurate results for the final analysis of interest. Second, since the final model is known, a tailor-made algorithm could be developed to produced the protected output, which optimizes the trade-off between accuracy and data protection.

However, there are some critical open questions that would need to be answered first if the approach is to be implemented in the future: First, the strategy requires a privacy budget to be assigned to each researcher. This budget would be used to protect the results that are to be released to the public. This leads to the dilemma outlined above that a decision needs to be made as to how the privacy budget should be distributed across researchers. Second, some decisions, such as which model should be used, will be made based on exploratory analyses working with the original data. This would need to be taken into account when releasing the results to avoid privacy leakage. Third, in the current debate on the reproducibility crisis, this approach seems to be a step in the wrong direction. If all of the privacy budget reserved for this research project were to be spent on protecting the final result, there would be no way to verify the results in an independent replication study. This problem could be mitigated by always reserving some privacy budget for replication purposes, of course, but this would negatively affect the accuracy of the publishable results, and deciding how much of the budget needed to be assigned for replication purposes would be challenging. Fourth, in several countries, RDCs are already struggling to satisfy the growing demand \citep{muller2019} and these demands might grow substantially if no public use microdata files were released anymore. Fifth, researchers might be tempted to select the final models to release and the algorithms for protecting them based on information from the data, which would violate the requirements of DP. Sixth, the onus of ensuring that the results can be published without violating confidentiality would be shifted from the data providing agency to the researcher. So far, the common practice is for staff at the RDC to check any output before the researcher can use it outside of the RDC facility. Now, the researchers would have to ensure that their final output was differentially private. This would raise two additional challenges: The researchers would have to be capable of developing algorithms that ensured DP for their output, which could be extremely challenging in practice as they would also need to account for the effects of all of the preprocessing and data cleaning steps on the final output. Furthermore, the staff at the RDC would have to verify that the approach taken by the researcher really ensured DP, which might be even harder in practice (see \citet{garfinkel2018} for a related point on finding qualified personnel and verifying that algorithms satisfy the requirements of DP in the context of the 2020 Decennial Census). Finally, the researcher needs to be trusted not to reveal any information about the data beyond the protected output. This might not seem like a major obstacle in most situations---after all, researchers are typically not interested in revealing any information beyond their research results and the common practice of imposing heavy fines if personally identifiable information is purposely released is a strong incentive to follow protocol. However, there will be situations in which researchers will be tempted to violate protocol. Imagine the following scenario: After spending several months of cleaning and preparing the data, extensive data exploration, careful model evaluations and several robustness checks, the researchers are finally ready to run the final model of interest and are overjoyed to see that the results strongly support the research hypothesis. Given the novelty and importance of the results, they are confident of being able to publish the findings in one of the leading journals in the field. All that is needed now is to obtain a DP version of the final model% (admittedly, this is already a simplifying assumption as major journals would also require proof for the robustness of the findings)
. Given the high stakes, the researchers decide to collaborate with some experts on DP to come up with an algorithm specifically tailored to the final model to ensure that the error introduced is as small as possible. However, when run on the data, the DP results are so different from the results based on the original data that they would contradict the null hypothesis. As the concept of differential privacy necessarily requires some randomness when moving from the original output to the protected output, an unlucky outcome of this nature could occur even for carefully designed algorithms and research output with low sensitivity. The question is whether the researchers would accept this outcome and silently bury all hopes of publishing the findings in a major journal. The problem arises because the researchers only have one shot at obtaining the final protected results. If they end up with an unlucky draw from the sanitization mechanism, they will have to accept the outcome. It seems questionable that researchers would accept such a lottery in practice.

There is, of course an analogy to sampling: even with carefully designed random samples, we can end up with a sample that is not representative of the population. However, with sampling we will never know whether this is the case or not. All we can do is to try to keep the sampling error as small as possible. This is fundamentally different from publishing results even if we know that the deviate substantially from the original results.

%Establishing such a system despite these problems might actually violate the privacy guarantees in the long run. As results would likely only be published if they do not deviate too much from the results based on the original data, an attacker would know that the bounds of the noise that was introduced are tighter than the bounds that would actually be required for the selected level of the privacy parameter $\varepsilon$.

One possible strategy for circumventing this problem would be to offer differentially private microdata to the accredited researchers for their exploratory analysis and to compute the final model on the original data only. Alternatively, verification or validation servers could be used. These servers take the analysis of interest and run it on both the protected and the original data, reporting back some metric for validity measuring how close the results from the protected data are to those that would have been obtained from the original data \citep{barrientos2018}. Still, %working only with noisy microdata would likely not be an option for most researchers. Furthermore,
some of the privacy budget would need to be used for generating the protected data, which would mean less budget for the accredited researchers.

\section{Understanding the privacy and accuracy achieved}\label{sec:priv_accuracy}
Part of the attractiveness of the DP approach lies in its intuitiveness. Differential privacy ensures that the influence that a single record can have on the reported output is strictly limited. This implies that the information that can be revealed about a single individual is also strictly limited. However, understanding the impacts on the accuracy of the results obtained is more difficult and even the interpretation of the level of protection that is provided for a fixed level of $\varepsilon$ is challenging.

\subsection{Impacts on accuracy}
Differential privacy has been criticised repeatedly for its negative impacts on the accuracy of the results obtained \citep{fienberg2010,bambauer2013}. Bounding the risks in a worst-case scenario considering all possible databases that differ in one record necessarily requires strong protection mechanisms. Nevertheless, there are many applications in which the approach can provide useful insights. Assessing the impacts in advance is typically only possible for relatively simple algorithms such as the Laplace mechanism, however. With these algorithms, the additional uncertainty that is introduced to protect the data can be quantified directly and can thus also be taken into account at the analysis stage (see the discussion on statistical inference below). Other mechanisms introduce the required randomness in a way that makes it difficult to assess the impacts on any subsequent analysis. The popular DP stochastic gradient descent method (SGDM)\citep{abadi2016}, for example, adds noise to the gradients of a neural network and also truncates their range. The effects on downstream analysis are difficult to assess analytically (see \citet{reiter2019} for a related discussion).

%The problem is magnified in the context of government agencies,  because as outlined in Section \ref{sec:no_query_response}, government agencies might have to provide differentially private microdata if they also aim at satisfying the data needs of academic researchers interested in running multivariate regression models on the data.
%Most of the algorithms that have been developed so far to generate synthetic data, such as the SGDM, fall into the second category of hard to quantify noise infusion techniques. Furthermore, the future analyses that will be performed on the differentially private microdata are typically unknown, making it impossible to anticipate the effects on accuracy and to optimize the algorithm to minimize these impacts.
Finally, postprocessing steps that are commonly employed to ensure that the protected data fulfill consistency requirements and do not contain implausible values such as negative age values (see \citet{barak2007, ding2011,hay2010,lee2015}, for example) also affect the accuracy of the results \citep{gong2020}. The optimization procedures that are commonly applied to find solutions close to the noisy result under consistency and non-negativity constraints again introduce biases in ways that are hard to anticipate and difficult to control for in any analysis using the protected data. In fact, \citet{Hawes2020} states that in one of the early test products released by the U.S. Census Bureau in preparation for the 2020 Decennial Census ``the postprocessing ... introduced far more error into the resulting data than came from the
differentially private noise used to protect privacy.'' Some early research attempting to address these problems is presented in  \citet{Zhu_2021} and \citet{gong2020}.

We would be remiss, of course, not to mention at this point that similar problems arise with most of the disclosure protection techniques that are currently employed at statistical agencies. Especially since details regarding the implemented procedures are commonly treated as confidential, it is often impossible in practice to account for the effects of these procedures when analyzing the data.

\subsection{Privacy guarantees}
While the general interpretation of the privacy guarantees offered by DP is intuitive, the risk implications of specific values of $\varepsilon$ are more difficult to understand. What are the actual risks to Bob if the probability of the algorithm producing a specific noisy answer to the query is, say, ten times more likely if he joins the database? Should he be worried? To put it in more concrete terms, consider the following simplified example. Assume a survey only asks about HIV status and plans to release the percentage of respondents that reported a positive status using an algorithm that satisfies the requirements of DP. If $\varepsilon=2.3$, Bob knows that the probability of the released value equaling any specific value $k$ if he participates in the survey, is never more than 10 times (since $e^{\varepsilon}\approx 10$) the probability of releasing the same value if he did not participate. This holds for any value of $k$, including the true value. However, how can he link this information to the risk of his own HIV status being revealed based on the released information? Understanding the impacts of changes in the value of $\varepsilon$ is even more difficult. How much of Bob's privacy is lost, if $\varepsilon$ is changed from 2.3 to 4.6? The probability ratio for observing a certain output with or without Bob increases from 10 to 100, but what does that mean for Bob? Obviously, understanding the direct risk implications of alternative protection strategies that are currently employed at government agencies is at least as difficult. Nevertheless, it is vitally important to understand the impacts of changes of $\varepsilon$ if a value of $\varepsilon$ should be found that addresses the trade-off between accuracy and privacy to an optimum degree.

Another challenge that arises when interpreting the privacy guarantees of DP is discussed in \citet{mcclure2012}. The authors compare the risks as measured by DP, with risk assessments based on more traditional statistical disclosure risk measures ``based on probabilities that intruders could learn information about data subjects given the released data and a set of assumptions about the intruder's knowledge and behavior.'' They find that the risks as measured by the statistical disclosure risk measures are small in some simulation settings, even for $\varepsilon=1,000$, and increasing $\varepsilon$ from, say, 1 to 10 sometimes has minor effects on the statistical disclosure risk measures (while substantially improving the analytical validity). These inconsistencies arise as DP is purposely data agnostic, and always assumes the worst-case scenario, whereas traditional disclosure risk measures strongly depend on the actual data that have been collected. The authors argue that risk measures should be estimated for the actual data and not including datasets that were not observed. They also propose some new risk measures that try to reconcile the two different perspectives.

When discussing the concept of DP with the interested public it is also important to emphasize that applying DP will not be enough to ensure that the data are protected. The level of protection will obviously depend on the selected value of $\varepsilon$ but also on the selected mechanism to achieve differential privacy.

Once again, a simplified example will help to illustrate this point. Assume a database contains five categorical variables: two of them are binary and three of them have three categories. The agency decides to use the geometric mechanism \citep{Ghosh:2009}, which is also the workhorse in the TopDown mechanism used for the 2020 Decennial Census (for the sake of simplicity we ignore any hierarchical data structures or postprocessing steps \citep{abowd2019TopDown}). The algorithm consists of three steps. In the first step, all variables are fully cross-classified, resulting in a contingency table with up to $2^2\cdot3^3=108$ cells (structural zeros, that is, implausible value combinations can be dropped). In the second step, random noise from a two-sided geometric distribution is added independently to each of the cells of the contingency tables. In the final step, the noisy table is turned back into privacy protected microdata to be released (potentially using some postprocessing to deal with negative counts in the noisy table). The agency decides to set $\varepsilon=8.6$, the value used in the U.S. Census Bureau's OnTheMap visualization tool. Assuming unbounded DP, this implies that the probability of any particular cell remaining unchanged, that is, of a value of zero being added to the true count, is $99.96\%$. Since noise is added to each cell independently, this implies that with a probability of more than $96.1\%$, the released data would match the original data in every single record (subject to random shuffling of the records). Thus, a potential attacker who recognized a record based on some of the attribute values could be very confident that the remaining, potentially sensitive attributes would still contain the true values for this record. This holds irrespective of how many records the database contains.

To be fair, these probabilities decrease quickly with the number of variables, and postprocessing steps will further reduce these probabilities%, and such large values of $\varepsilon$ are generally not recommended
. The point is that it is important to consider the context and to be aware that DP does not automatically guarantee an acceptable level of protection unless a sufficiently small value of $\varepsilon$ is selected.

\section{Differential privacy in the survey context}\label{sec:surveys}
Most information obtained by statistical agencies is collected through surveys which typically comprise only a small sample from the underlying population of interest. Thus, it will be crucial to understand the impacts of DP on survey data. Some open questions in this context will be discussed in this section.

\subsection{Incentives to share the data}
In all successful deployments of DP in the industry context, the data providers have strong incentives to provide their data, as they derive some benefit from the services offered by these companies in exchange for the data. Take the example of the deployment of DP for Google Chrome \citep{erlingsson2014}: The benefits of having access to a convenient internet browser make it seem acceptable for some information about my browsing behavior to be shared with Google. It is the price that I pay for being able to use this service. Furthermore, the well-known privacy paradox states that although users express concerns about their privacy, their behavior does not reflect these concerns \citep{taddicken2014}. This paradox has been confirmed in several independent studies: the desire to use the products seems to outweigh any privacy concerns.

In the survey context the situation is different. Although there are obvious benefits to society that are shown by the insights obtained from the collected data, the direct benefits for the survey respondent are less obvious. While many surveys offer respondents a small recompense for participating in the survey, the amounts offered are low to avoid introducing bias into the collected data because the incentive might work better for some subgroups than for others. Partly because of these limited benefits, nonresponse rates have been constantly increasing over the years. While response rates above sixty percent were still achievable in 1997, these rates dropped to $22$ percent by 2012 according to a study conducted by the PEW Research Center \citep{kohut2012}.

At first sight, offering strong formal privacy guarantees to respondents will improve the quality of the collected data, as more respondents might be willing to participate and to respond truthfully if they know that confidentiality is guaranteed. However, this only holds if the respondents fully understand and trust the protection mechanisms. Previous experience shows that this is not always the case. Randomized response, as proposed by \citet{warner1965}, which coincidentally happens to be the earliest mechanism satisfying the requirements of DP, protects answers to sensitive questions by introducing randomness into the response process. In its basic form, the respondent flips a coin, and depending on the result of the coin flip, either answers the sensitive question or provides an answer to a completely different question that is non-sensitive. As only the respondent knows whether the reported answer is the answer to the sensitive or the non-sensitive question, her sensitive information is protected. However, several studies have shown that introducing the concept of randomized response in a survey does not increase response rates or the likelihood of respondents answering truthfully \citep{edgell1982,landsheer1999,coutts2011,kirchner2015}. Respondents do not seem to trust the concept enough to reveal truthful answers (see \citet{oberski2020} for a similar argument).
Thus, as discussed in the previous section, in practice it will not be sufficient to understand the actual privacy guarantees of differential privacy. Data collecting agencies will also have to take into account the perceived level of privacy protection and the implications this might have for the willingness to participate in surveys in the future.

On the other hand, guaranteeing DP might undermine the willingness to participate. Survey vendors often try to motivate potential participants by emphasizing the relevance of the study and the important insights that might be obtained from the collected data. However, DP sends a different signal, basically telling the respondent: no matter what information you provide, we will make sure it is irrelevant for the final findings. The results will be more or less the same even if you do not participate. This message might destroy one of the few remaining incentives for survey participation: the feeling that the time spent answering a long and boring questionnaire is well spent as the answers provided will make an important contribution to helping researchers to understand better and potentially improve society (see \citet{kreuter2019} for a similar argument).

\subsection{Complications regarding inference validity}\label{sec:valid_inference}
With survey data, the goal is often to make inferences regarding an underlying (finite) population, that is, the results based on the survey data are treated as estimates for the true values in the population. Furthermore, there is often a major interest in identifying causal effects. To be able to achieve these goals, it is essential to quantify the uncertainty in the estimates obtained from the survey data, and sophisticated methods have been developed in the survey statistics literature to do so while accounting for complex sampling designs and nonresponse adjustments (see \citet{sarndal2003}, for example).

One key challenge which needs to be addressed when dealing with DP for survey data, is how to obtain statistically valid inferences from the protected data, that is, how to take the extra uncertainty due to the protection mechanism into account. Research in this area is still limited. While it is straightforward to quantify the extra uncertainty for simple algorithms such as the Laplace mechanism, which simply adds noise to the generated output, it is challenging to measure the uncertainty for more complex algorithms or analyses that are functions of the noisy output. Furthermore, some algorithms introduce data-dependent noise or systematic bias, both of which are difficult to control for, especially for multivariate statistics. Finally, postprocessing steps, which are typically performed to ensure consistency and to avoid implausible values, %such as negative cell counts in a frequency table
further affect the final estimates in complex ways, invalidating potential adjustments even for simple algorithms.

\citet{gong2020transparent} presents a conceptual framework to address these challenges. The basic idea of this framework is to integrate any privacy procedures into downstream analyses by modeling the protection mechanism and integrating over the unknown, confidential data (see \citet{little1993} for a related idea). The paper includes an illustrative example based on linear regression in the context of differential privacy. Nonetheless, modeling the protection mechanism can be challenging in practice, especially if postprocessing steps need to be taken into account.

Nevertheless, it is important again to emphasize that it is often similarly (if not more) challenging to properly account for the impacts of traditionally employed protection strategies such as swapping, since the privacy parameters are considered confidential, although there are examples where statistical agencies provide margins of error that try to account for the extra uncertainty introduced in order to ensure that the output is sufficiently protected \citep{mckinney2020}.

\subsection{No secrecy of the sample?}
Common sense would suggest that sampling offers additional protection from certain types of risks, as a potential attacker does not know whether the target is part of the sample or not. This suggestion has been formalized in the DP literature for certain sampling types such as simple random sampling with or without replacement or Poisson sampling \citep{Balle2018}, showing that these sampling designs can indeed lead to privacy amplification. With simple random sampling without replacement, for example, a $\varepsilon$-DP output computed on a sample selected with sampling rate $r$ actually offers a privacy guarantee of approximately $r\varepsilon$ (for small values of $\varepsilon$).  %For example, in a blog post, Adam Smith (\citeyear{Smith2009}) illustrated that given a mechanism offering a privacy guarantee of $\varepsilon_1$ if run on the entire population and a sampling design based on simple random sampling with replacement with sampling rate $r$, the privacy guarantee provided by running the mechanism on the sample instead of the population is $\varepsilon_2=r\varepsilon_1$.

Quantifying the privacy amplification offered through sampling would allow agencies to either report more accurate statistics for a given privacy budget or offer a higher level of data protection to the respondents. However, the sampling designs employed by government agencies are typically far more complex than the ones studied so far. Stratified, cluster, and PPS sampling are commonly applied to improve the efficiency and/or reduce the costs of data collection. Furthermore, the final sample is typically drawn in multiple stages, with different sampling strategies being combined at the various levels.

Initial research indicates that complex sampling designs can actually lead to privacy degradation instead of amplification. \citet{bun2020} find that stratified sampling designs can negatively affect the privacy guarantees, and amplification will generally be negligible for cluster sampling. %However, they also find that privacy amplification can be retained for stratified sampling using proportional allocation, if random rounding is used instead of conventional rounding when determining the final sample sizes in each stratum.
One important aim for future research is to gain a better understanding of the impacts of (multistage) sampling designs on the final privacy guarantees that can be offered. It might also be desirable in the long run to already take the privacy implications into account when designing a survey. A more efficient sampling design might not be helpful if much more noise needs to be added in the end to protect the output sufficiently.

\subsection{Dealing with survey weights}
One implication of using complex sampling designs as discussed in the previous section is that the probability of being selected into the survey typically differs between the units. To simplify the analysis for the users, statistical agencies provide survey weights, which are computed as the inverse of the probability of being included in the survey. In survey weighted analyses, the contribution of each unit to the final estimate is typically weighted by its survey weight. This has strong implications in the DP context. To illustrate, we could consider a counting query, which, as discussed previously, has low sensitivity, since changing one record in the database will have small impacts on the final counts. However, for survey weighted counts, the sensitivity will depend on the supremum of the weights, that is, the maximum value of the weights over all possible neighboring datasets. Since the amount of noise that needs to be infused to ensure DP is proportional to the sensitivity, much more noise will be required to protect the survey weighted estimate properly (see \citet{reiter2019} for a related argument).
%To illustrate the challenges that can arise with survey weighted estimates, let us assume the analyst is interested in estimating the number of females aged between 40 and 45 in the population. Remember that with differential privacy the amount of noise that needs to be infused is typically dependent on the sensitivity of the output, that is, how much the output would change if we change the value of a single record in the data. The sensitivity of our statistic of interest would be 1, if we were only interested in counting the number of females between 40 and 45 in our data as the total count can only change by one if we are only allowed to change one record in the data. However, to estimate this number for the population based on the survey data, the Horwitz-Thompson estimator \citep{Horvitz1952} would typically be used. Let $Z_i$ be an indicator variable that equals one if unit $i$ is female and between 40 and 45. The population estimate would be given by $\hat{t}=\sum Z_iw_i$, where $w_i$ is the survey weight for unit $i$ and the summation is over all the units in the sample. Obviously, the sensitivity of this statistic is the supremum of the survey weights. This illustrates that the sensitivity of survey weighted estimates is typically larger than the sensitivity if the full target population is available

However, the more challenging problem is to identify the supremum of the weights. Even if we only consider the design weights, that is, the weights that account for the different probabilities of selection because of the sampling design, identifying the supremum can be an obstacle. Depending on the desired privacy guarantees, we would need to consider how changing the values of one record in the population can change the probabilities of selection of all the other units. This can be challenging for data-dependent sampling designs such as stratified sampling using Neyman allocation, where the probabilities of selection depend on the standard deviations of the design variable in each stratum.

Identifying the supremum is further complicated by the fact that the initial weights are typically adjusted to account for nonresponse and other deficiencies, such as undercoverage in the sampling frame. These adjustment steps typically increase the variability of the design weights, further inflating the sensitivity of the estimate. Furthermore, given that commonly used adjustment steps such as nonresponse adjustments or calibration depend on the collected data, it will become even more difficult to identify the supremum of the weights. %Finally, as discussed below it will be important to ensure that these adjustment steps do not affect the privacy guarantees as the resulting weights might reveal some information regarding the underlying data.

\subsection{Understanding the impacts of preprocessing}
Statistical agencies aim at collecting samples that are representative of the target population and employ various methods to ensure that valid inferences can be obtained regarding the population of interest. Before any analyses are run, the raw data are cleaned and updated in several preprocessing steps. Unit nonresponse is addressed by adjusting the survey weights, item nonresponse is often addressed by imputing missing values, and implausible values are corrected using various editing and imputation procedures. Currently, the understanding how the typical data production process affects the privacy guarantees or how the process could be adjusted to satisfy DP is limited \citep{reiter2019}.

Assuming the goal is to release differentially private microdata, one key question is whether these preprocessing steps should be carried out before or after generating the protected data. There are arguments for both approaches. Data editing, for example, mostly involves finding and adjusting implausible outliers by establishing editing rules that raise a flag whenever a record violates a rule. Since editing can be seen as a form of clamping or truncation of the raw data, which typically helps to reduce the sensitivity of potential queries to be run on the data, it might actually improve the accuracy of analyses that respect privacy if editing rules were enforced before protecting the data.

Intuitively, having imputed values in the data could also increase the level of privacy, as these values are only estimates for the true values and thus already introduce an extra level of uncertainty. However, at their core, all imputation strategies rely on models which are estimated using the observed data, and they can therefore increase privacy leakage. The extent to which sensitive information is leaked depends on the imputation strategy used. Early research~\citep{clifton2019} found, for example, that hot-deck imputation, popular amongst statistical agencies, is problematic as it imputes missing values by transferring the values of similar records that are fully observed (and is thus highly sensitive to changes in the data). This would be an argument for moving away from donor-based imputation methods if privacy considerations are also taken into account.

An alternative approach would be to start by generating differentially private data first and then imputing based on the generated data. As DP is immune to postprocessing, the imputation routines would not require any extra privacy budget as long as they only rely on the DP data. This would also offer the option of releasing DP microdata still containing missing values to allow researchers to decide themselves which strategy they consider most suitable for dealing with the missing data. However, this raises the question of how to deal with missing values when enforcing differential privacy.

Similar questions arise for the weighting adjustments that are commonly applied to control for unit nonresponse and other deficiencies, such as undercoverage in the sampling frame. If unit-nonresponse adjustments and calibration techniques---which ensure that the weighted survey estimates exactly match certain values known for the population---are used prior to protecting the data, it will be important to understand their effects on the sensitivity of the data. If the data are protected first, this would require all the information which is used for the weighting adjustments to be protected additionally. For nonresponse weighting this would include the design variables from the sampling frame; for calibration this would imply that only noisy benchmarks from the population could be used, unless these benchmarks can be treated as public knowledge that does not need to be protected. %Adopting this strategy raises challenging questions how to account for these extra measures when trying to make inferences regarding the population of interest.

 %All these processes will affect the privacy guarantees in different ways and research in this area has been limited so far.

\section{Setting the value of $\varepsilon$}\label{sec:setting_eps}
Setting the value for the privacy parameter $\varepsilon$ is arguably the most difficult decision in any deployment of DP in practice. This is partly due to the difficulty of understanding the implications of different values of $\varepsilon$ for privacy and accuracy, as discussed in Section \ref{sec:priv_accuracy}, but also because the optimal value is inherently a social choice which begs the question: how much privacy am I willing to give up for more accurate results? \citet{schmutte2019} approach the problem from an economic perspective, arguing that an optimal solution should ``operate where
the marginal cost of increasing privacy equals the marginal benefit.'' The authors provide several examples of how this could be done in practice, but also acknowledge the challenges in measuring these costs and benefits. Nevertheless, there are several aspects, which make finding the optimal value for $\varepsilon$ specifically challenging for government agencies.

\subsection{Government agencies often act as intermediates}
From an economic perspective, finding an optimal value of $\varepsilon$ seems to be straightforward at first sight, especially in the industry context, as only two parties with seemingly simple utility functions are involved. The data provider will aim at minimizing the amount of privacy leakage while still getting the service offered by the company. The data recipient will try to maximize the privacy leakage to obtain more accurate data, while ensuring the data provider is still willing to participate. To illustrate, we can consider an internet search engine provider. The company running the engine would ideally like to store all of the information entered by the user (maximize privacy leakage) as the information can be used to generate revenue from targeted advertising. The user, that is, the data provider, ideally does not want her information to be used for other purposes, but is willing to give up some privacy to be able to use the service. In this stylized setting, economic theory tells us that we can expect a market equilibrium at the optimal value of $\varepsilon$.

In practice, however, measuring the costs and benefits of sharing private information both on the personal level and for society is much more nuanced (see \citet{acquisti2016} for an excellent review). Challenges such as limited awareness of the consequences of data sharing, information asymmetries, and limited market access  ``raise questions regarding individuals’ abilities, as rational consumers, to optimally navigate privacy trade-offs,'' which ultimately raises concerns about whether there are ``privacy `equilibria´ that benefit both data holders and data subjects'' \citep{acquisti2016}.
%Leaving aside all the challenges in understanding $\varepsilon$, information asymmetries, and the privacy paradox,

The situation is even more complex in the government context, since the data collecting agencies are often not the end users of the collected data. Data collected by national statistical agencies such as the U.S. Census Bureau, for example, are used by various stakeholders, including politicians, journalists, and social scientists. The agencies not only face the dilemma that they need to anticipate the utility functions of all of these stakeholders, but that these utility functions will likely also vary considerably between the stakeholders. Furthermore, as discussed above, the agency cannot anticipate all of the future analyses that users will be interested in, making it impossible to evaluate the impacts on accuracy. %In a worst case scenario, the agency will settle on a value of $\varepsilon$ that data providers consider insufficient to protect the data while the potential users of the data are no longer willing to work with the data as they feel that too much accuracy has been compromised.

\subsection{Low values of $\varepsilon$ potentially harmful for the data provider}
There is another aspect that makes finding the optimal value for $\varepsilon$ more challenging in the government context. Once the data have been shared, the party benefitting most from the data in the industry context is the company. Further usage of the data will typically only have negative effects for the data provider if, for example, the company sells the data to a data broker (see \citet{acquisti2016} for some counterexamples, however). Thus, the primary goal of the provider will always be to keep the privacy leakage as small as possible.

This could be different in the government context, where data with low accuracy might be at least as harmful as data with low levels of privacy. To illustrate, we can look at the results of one of the early test runs of the U.S. Census Bureau's TopDown algorithm on the 2010 Decennial Census. As illustrated in a New York Times article by \citet{NYT2020}, the version of the algorithm used in the test run introduced a systematic downward bias in the counts of Native Americans on reservations. Since, according to the Census Bureau, the counts from the next decennial census will be used ``to inform the allocation of hundreds of billions in federal funding'' \citep{Census2020}, it is obvious that any underestimation in the counts will have direct negative impacts for the subjects involved. The problem has been fixed in the TopDown algorithm in the meantime and it is important to re-emphasize that the primary focus of early demonstration products released by the Census Bureau was on privacy and not accuracy. %The privacy parameters that are used for the actual data releases based on the Decennial Census 2020 are substantially higher than those used in the test runs implying a shift from a pure focus on privacy protection to also guaranteeing a high level of accuracy \citep{Census2021privacy}.
Nevertheless, this example illustrates that the utility function of the data providers might be more complex, as low levels of accuracy can have strong impacts on their utility. Arguably, there will likely be data providers in this context that would prefer to give up most of their privacy if they would otherwise risk receiving less government funding. Obviously, similar situations would arise in other scenarios---if government money were spent on health care measures based on DP results or poverty rates were computed using DP data, for example.

\subsection{The optimal value for society}
Finding the optimal value of $\varepsilon$ is difficult for various reasons. Obviously, different respondents will have different views on privacy. Healthy respondents, for example, will typically be much less concerned regarding questions about their health status than respondents suffering from a rare disease. However, how do we aggregate across these individual preferences to find the optimal value for society? Would it be ethically defensible if the released data were still accurate enough to allow health insurance companies to better target their premiums? This might beof benefit to the majority of respondents, as their premiums would decrease, but some respondents might have to pay the price of losing their insurance. Thus, simply maximizing the utility across respondents would not be an option. The problem is magnified by the fact that deciding not to participate in the survey would not save the respondent from these types of negative consequences.

\section{The road forward}
Despite its critical tone, this article should not be read as a general critique of DP, nor should it by any means imply that the concept is generally unsuitable for the government context. Being able to offer formal privacy guarantees would be a major leap forward regarding how statistical agencies value the privacy of their respondents.% and the reconstruction attacks conducted by the U.S. Census Bureau on some of their own previously released data products \citep{abowd2019} clearly illustrated that the protection methods that are still popular among statistical agencies around the world are no longer sufficient to adequately protect the data.

The aim of the article is to raise the awareness that adopting the concept of differential privacy at government agencies raises many challenges and questions that did not have to be addressed in previous deployments. The hope is to stimulate interdisciplinary research to address these challenges. To be fair, some challenges cannot be solved easily. It will always be true, for example, that the size of the database will be relatively small in the survey context. However, the increasing popularity of administrative records and found data for statistical purposes might mitigate some of these problems. Morover, little attention has been given so far to the performance of algorithms if run on small samples. Substantial gains in terms of accuracy might still be possible in practice if the focus shifts to developing optimal algorithms for fixed sample sizes.

Other challenges could be addressed by increased collaborations between disciplines. The vast majority of papers published on DP to date have been authored by computer scientists. Since topics such as statistical inference, complex sampling designs, or nonresponse adjustment typically only play a minor role in this field, it is not surprising that research in this area has so far been limited. While some papers that have appeared in recent years also look at the problem from a statistical perspective (\citet{wasserman2010, awan2020, karwa2016, bowen2020}), more collaborations between computer scientists and (survey) statisticians would help address many of the questions raised in Section \ref{sec:surveys} of this paper.

Finally, collaborations need to be expanded further when trying to identify the value of $\varepsilon$ that is optimal for society. Economists, social scientists, psychologists, experts on data ethics, and survey methodologists could all make invaluable contributions to a better understanding of individual feelings about privacy, identify models that accurately describe the social welfare function as discussed in \citet{schmutte2019}, and understand social behavior in response to threats to privacy, for instance. The successful collaborations between computer scientists and legal experts to study the legal implications of DP \citep{altman2015,nissim2017,nissim2018} already illustrate the knowledge gains that are possible from such collaborations.

Differential privacy is currently at a critical transition stage, moving from an attractive but purely theoretical concept to becoming the de facto standard that serves as a benchmark for any research on data privacy (for good or ill, it is almost impossible these days to publish any research that does not meet the DP requirements without detailed explanations as to why DP methods were not considered). Whether the concept will also be adopted for the data dissemination strategies of government agencies in practice will depend on answers and solutions to the questions and challenges raised in this article being found. The theoretical properties of DP have been studied extensively over the last decade. Now is the time to gain a better understanding of the implications of adopting the idea in practice, and this will require a joint effort from different disciplines as many of the problems arising in practice are clearly outside of the realm of theoretical computer science.

\section*{Acknowledgements}
The work on this project was funded in part by US Census Bureau cooperative agreement CB20ADR0160001. The opinions, findings, conclusions and recommendations expressed herein are those of the author and do not necessarily reflect the views of the US Census Bureau. I am  grateful for helpful feedback on earlier versions of this paper from (in no particular order) Audra McMillan, Philip Leclerc, Michael Freiman, and Mark Fleischer.

\bibliographystyle{natbib}

\bibliography{mybib}

\begin{thebibliography}{}

\bibitem[Abadi \emph{et~al.}(2016)Abadi, Chu, Goodfellow, McMahan, Mironov,
  Talwar, and Zhang]{abadi2016}
Abadi, M., Chu, A., Goodfellow, I., McMahan, H.~B., Mironov, I., Talwar, K.,
  and Zhang, L. (2016).
\newblock Deep learning with differential privacy.
\newblock In \emph{Proceedings of the 2016 ACM SIGSAC Conference on Computer
  and Communications Security},  308--318.

\bibitem[Abowd \emph{et~al.}(2019)Abowd, Ashmead, Simson, Kifer, Leclerc,
  Machanavajjhala, Moran, Sexton, and Zhuravlev]{abowd2019TopDown}
Abowd, J., Ashmead, R., Simson, G., Kifer, D., Leclerc, P., Machanavajjhala,
  A., Moran, B., Sexton, W., and Zhuravlev, P. (2019).
\newblock Census {T}op{D}own algorithm: Differentially private data,
  incremental schemas, and consistency with public knowledge.
\newblock
  \url{https://github.com/uscensusbureau/census2020-das-2010ddp/blob/master/doc/20191020_1843_Consistency_for_Large_Scale_Differentially_Private_Histograms.pdf}.

\bibitem[Abowd(2018)]{abowd2018}
Abowd, J.~M. (2018).
\newblock The {U.S.} {C}ensus {B}ureau adopts differential privacy.
\newblock In \emph{Proceedings of the 24th ACM SIGKDD International Conference
  on Knowledge Discovery \& Data Mining},  2867.

\bibitem[Abowd and Schmutte(2019)]{schmutte2019}
Abowd, J.~M. and Schmutte, I.~M. (2019).
\newblock An economic analysis of privacy protection and statistical accuracy
  as social choices.
\newblock \emph{American Economic Review} \textbf{109}, 1, 171--202.

\bibitem[Acquisti \emph{et~al.}(2016)Acquisti, Taylor, and
  Wagman]{acquisti2016}
Acquisti, A., Taylor, C., and Wagman, L. (2016).
\newblock The economics of privacy.
\newblock \emph{Journal of Economic Literature} \textbf{54}, 2, 442--492.

\bibitem[Altman \emph{et~al.}(2015)Altman, Wood, O’Brien, Vadhan, and
  Gasser]{altman2015}
Altman, M., Wood, A., O’Brien, D.~R., Vadhan, S., and Gasser, U. (2015).
\newblock Towards a modern approach to privacy-aware government data releases.
\newblock \emph{Berkeley Technology Law Journal} \textbf{30}, 3, 1967--2072.

\bibitem[{Apple's Differential Privacy Team}(2017)]{adp2017}
{Apple's Differential Privacy Team} (2017).
\newblock Learning with privacy at scale.
\newblock \emph{Apple Machine Learning Journal} \textbf{1}, 8.

\bibitem[Awan and Slavkovi{\'c}(2021)]{awan2020}
Awan, J. and Slavkovi{\'c}, A. (2021).
\newblock Structure and sensitivity in differential privacy: Comparing k-norm
  mechanisms.
\newblock \emph{Journal of the American Statistical Association} \textbf{116},
  534, 935--954.

\bibitem[Bailie and Chien(2019)]{bailie2019}
Bailie, J. and Chien, C.-H. (2019).
\newblock {ABS} perturbation methodology through the lens of differential
  privacy.
\newblock In \emph{UNECE work session on statistical data confidentiality}.

\bibitem[Balle \emph{et~al.}(2018)Balle, Barthe, and Gaboardi]{Balle2018}
Balle, B., Barthe, G., and Gaboardi, M. (2018).
\newblock Privacy amplification by subsampling: Tight analyses via couplings
  and divergences.
\newblock In \emph{Advances in Neural Information Processing Systems 31: Annual
  Conference on Neural Information Processing Systems 2018, NeurIPS 2018, 3-8
  December 2018, Montr{\'{e}}al, Canada},  6280--6290.

\bibitem[Bambauer \emph{et~al.}(2013)Bambauer, Muralidhar, and
  Sarathy]{bambauer2013}
Bambauer, J., Muralidhar, K., and Sarathy, R. (2013).
\newblock Fool's gold: {A}n illustrated critique of differential privacy.
\newblock \emph{Vand. J. Ent. \& Tech. L.} \textbf{16}, 701.

\bibitem[Barak \emph{et~al.}(2007)Barak, Chaudhuri, Dwork, Kale, McSherry, and
  Talwar]{barak2007}
Barak, B., Chaudhuri, K., Dwork, C., Kale, S., McSherry, F., and Talwar, K.
  (2007).
\newblock Privacy, accuracy, and consistency too: {A} holistic solution to
  contingency table release.
\newblock In \emph{Proceedings of the Twenty-Sixth ACM SIGMOD-SIGACT-SIGART
  Symposium on Principles of Database Systems},  273--282.

\bibitem[Barrientos \emph{et~al.}(2018)Barrientos, Bolton, Balmat, Reiter,
  de~Figueiredo, Machanavajjhala, Chen, Kneifel, and DeLong]{barrientos2018}
Barrientos, A.~F., Bolton, A., Balmat, T., Reiter, J.~P., de~Figueiredo, J.~M.,
  Machanavajjhala, A., Chen, Y., Kneifel, C., and DeLong, M. (2018).
\newblock Providing access to confidential research data through synthesis and
  verification: An application to data on employees of the {U.S.} federal
  government.
\newblock \emph{The Annals of Applied Statistics} \textbf{12}, 2, 1124--1156.

\bibitem[Bowen and Liu(2020)]{bowen2020}
Bowen, C.~M. and Liu, F. (2020).
\newblock Comparative study of differentially private data synthesis methods.
\newblock \emph{Statistical Science} \textbf{35}, 2, 280--307.

\bibitem[boyd(2019)]{boyd2019}
boyd, D. (2019).
\newblock Differential privacy in the 2020 {D}ecennial {C}ensus and the
  implications for available data products.
\newblock Available at SSRN: \url{https://ssrn.com/abstract=3416572}.

\bibitem[Bun \emph{et~al.}(2020)Bun, Drechsler, Gaboardi, and
  McMillan]{bun2020}
Bun, M., Drechsler, J., Gaboardi, M., and McMillan, A. (2020).
\newblock Controlling privacy loss in survey sampling.
\newblock \emph{arXiv preprint arXiv:2007.12674} .

\bibitem[Clifton \emph{et~al.}(2020)Clifton, Hanson, Merrill, and
  Merrill]{clifton2019}
Clifton, C., Hanson, Eric, J., Merrill, K., and Merrill, S. (2020).
\newblock Differentially private k-nearest neighbor missing data imputation.
\newblock Working Paper cedWP-2020-070, U.S. Census Bureau.

\bibitem[Coutts and Jann(2011)]{coutts2011}
Coutts, E. and Jann, B. (2011).
\newblock Sensitive questions in online surveys: Experimental results for the
  randomized response technique ({RRT}) and the unmatched count technique
  ({UCT}).
\newblock \emph{Sociological Methods \& Research} \textbf{40}, 1, 169--193.

\bibitem[Ding \emph{et~al.}(2017)Ding, Kulkarni, and Yekhanin]{ding2017}
Ding, B., Kulkarni, J., and Yekhanin, S. (2017).
\newblock Collecting telemetry data privately.
\newblock In \emph{Advances in Neural Information Processing Systems},
  3571--3580.

\bibitem[Ding \emph{et~al.}(2011)Ding, Winslett, Han, and Li]{ding2011}
Ding, B., Winslett, M., Han, J., and Li, Z. (2011).
\newblock Differentially private data cubes: {O}ptimizing noise sources and
  consistency.
\newblock In \emph{Proceedings of the 2011 ACM SIGMOD International Conference
  on Management of Data},  217--228.

\bibitem[Dinur and Nissim(2003)]{dinur2003}
Dinur, I. and Nissim, K. (2003).
\newblock Revealing information while preserving privacy.
\newblock In \emph{Proceedings of the Twenty-Second ACM SIGMOD-SIGACT-SIGART
  Symposium on Principles of Database Systems},  202--210.

\bibitem[Domingo{-}Ferrer \emph{et~al.}(2020)Domingo{-}Ferrer, S{\'{a}}nchez,
  and Blanco{-}Justicia]{josep2020}
Domingo{-}Ferrer, J., S{\'{a}}nchez, D., and Blanco{-}Justicia, A. (2020).
\newblock The limits of differential privacy (and its misuse in data release
  and machine learning).
\newblock \emph{CoRR} \textbf{abs/2011.02352}.

\bibitem[Dwork \emph{et~al.}(2006)Dwork, Mcsherry, Nissim, and
  Smith]{Dwork2006}
Dwork, C., Mcsherry, F., Nissim, K., and Smith, A. (2006).
\newblock Calibrating noise to sensitivity in private data analysis.
\newblock In \emph{Proceedings of the 3rd Theory of Cryptography Conference},
  265--284.

\bibitem[Dwork and Roth(2014)]{dwork2014}
Dwork, C. and Roth, A. (2014).
\newblock The algorithmic foundations of differential privacy.
\newblock \emph{Foundations and Trends in Theoretical Computer Science}
  \textbf{9}, 3-4, 211--407.

\bibitem[Edgell \emph{et~al.}(1982)Edgell, Himmelfarb, and Duchan]{edgell1982}
Edgell, S.~E., Himmelfarb, S., and Duchan, K.~L. (1982).
\newblock Validity of forced responses in a randomized response model.
\newblock \emph{Sociological Methods \& Research} \textbf{11}, 1, 89--100.

\bibitem[Erlingsson \emph{et~al.}(2014)Erlingsson, Pihur, and
  Korolova]{erlingsson2014}
Erlingsson, {\'U}., Pihur, V., and Korolova, A. (2014).
\newblock Rappor: Randomized aggregatable privacy-preserving ordinal response.
\newblock In \emph{Proceedings of the 2014 ACM SIGSAC Conference on Computer
  and Communications Security},  1054--1067.

\bibitem[Fienberg \emph{et~al.}(2010)Fienberg, Rinaldo, and Yang]{fienberg2010}
Fienberg, S.~E., Rinaldo, A., and Yang, X. (2010).
\newblock Differential privacy and the risk-utility tradeoff for
  multi-dimensional contingency tables.
\newblock In \emph{International Conference on Privacy in Statistical
  Databases},  187--199. Springer.

\bibitem[Foote \emph{et~al.}(2019)Foote, Machanavajjhala, and
  McKinney]{Foote2019}
Foote, A.~D., Machanavajjhala, A., and McKinney, K. (2019).
\newblock Releasing earnings distributions using differential privacy:
  Disclosure avoidance system for post-secondary employment outcomes (pseo).
\newblock \emph{Journal of Privacy and Confidentiality} \textbf{9}, 2.

\bibitem[Garfinkel \emph{et~al.}(2019)Garfinkel, Abowd, and
  Martindale]{garfinkel2019}
Garfinkel, S., Abowd, J.~M., and Martindale, C. (2019).
\newblock Understanding database reconstruction attacks on public data.
\newblock \emph{Communications of the ACM} \textbf{62}, 3, 46--53.

\bibitem[Garfinkel \emph{et~al.}(2018)Garfinkel, Abowd, and
  Powazek]{garfinkel2018}
Garfinkel, S.~L., Abowd, J.~M., and Powazek, S. (2018).
\newblock Issues encountered deploying differential privacy.
\newblock In \emph{Proceedings of the 2018 Workshop on Privacy in the
  Electronic Society},  133--137.

\bibitem[Ghosh \emph{et~al.}(2009)Ghosh, Roughgarden, and
  Sundararajan]{Ghosh:2009}
Ghosh, A., Roughgarden, T., and Sundararajan, M. (2009).
\newblock Universally utility-maximizing privacy mechanisms.
\newblock In \emph{Proceedings of the 41st Annual ACM Symposium on Theory of
  Computing}, STOC '09,  351--360, New York. ACM.

\bibitem[Gong(2020)]{gong2020transparent}
Gong, R. (2020).
\newblock Transparent privacy is principled privacy.
\newblock \emph{arXiv preprint arXiv:2006.08522} .

\bibitem[Gong and Meng(2020)]{gong2020}
Gong, R. and Meng, X.-L. (2020).
\newblock Congenial differential privacy under mandated disclosure.
\newblock In \emph{Proceedings of the 2020 ACM-IMS on Foundations of Data
  Science Conference},  59--70.

\bibitem[Hallowell and Rector(2020)]{Maine_letter2020}
Hallowell, A. and Rector, A. (2020).
\newblock Maine state economist letter to census on differential privacy.
\newblock Available at:
  \url{https://sdcclearinghouse.com/2020/02/27/maine-state-economist-letter-to-census-on-differential-privacy/}.

\bibitem[Hawes(2020)]{Hawes2020}
Hawes, M.~B. (2020).
\newblock Implementing differential privacy: Seven lessons from the 2020
  {U}nited {S}tates {C}ensus.
\newblock \emph{Harvard Data Science Review} \textbf{2}, 2.
\newblock https://hdsr.mitpress.mit.edu/pub/dgg03vo6.

\bibitem[Hay \emph{et~al.}(2010)Hay, Rastogi, Miklau, and Suciu]{hay2010}
Hay, M., Rastogi, V., Miklau, G., and Suciu, D. (2010).
\newblock Boosting the accuracy of differentially-private queries through
  consistency.
\newblock In \emph{36th International Conference on Very Large Databases
  (VLDB)}. Citeseer.

\bibitem[{IPUMS USA}(2021)]{IPUMS2021}
{IPUMS USA} (2021).
\newblock 1940 bibliography.
\newblock \url{https://usa.ipums.org/usa/1940bibliography.shtml}.

\bibitem[Karwa and Slavkovi{\'c}(2016)]{karwa2016}
Karwa, V. and Slavkovi{\'c}, A. (2016).
\newblock Inference using noisy degrees: Differentially private $\beta$-model
  and synthetic graphs.
\newblock \emph{The Annals of Statistics} \textbf{44}, 1, 87--112.

\bibitem[Kirchner(2015)]{kirchner2015}
Kirchner, A. (2015).
\newblock Validating sensitive questions: A comparison of survey and register
  data.
\newblock \emph{Journal of Official Statistics} \textbf{31}, 1, 31--59.

\bibitem[Kohut \emph{et~al.}(2012)Kohut, Keeter, Doherty, Dimock, and
  Christian]{kohut2012}
Kohut, A., Keeter, S., Doherty, C., Dimock, M., and Christian, L. (2012).
\newblock Assessing the representativeness of public opinion surveys.
\newblock \emph{Washington, DC: Pew Research Center} .

\bibitem[Kreuter(2019)]{kreuter2019}
Kreuter, F. (2019).
\newblock The social survey statistician's perspective.
\newblock Online video available at:
  https://simons.berkeley.edu/talks/perspective-1.

\bibitem[Landsheer \emph{et~al.}(1999)Landsheer, Van Der~Heijden, and
  Van~Gils]{landsheer1999}
Landsheer, J.~A., Van Der~Heijden, P., and Van~Gils, G. (1999).
\newblock Trust and understanding, two psychological aspects of randomized
  response.
\newblock \emph{Quality and Quantity} \textbf{33}, 1, 1--12.

\bibitem[Lee \emph{et~al.}(2015)Lee, Wang, and Kifer]{lee2015}
Lee, J., Wang, Y., and Kifer, D. (2015).
\newblock Maximum likelihood postprocessing for differential privacy under
  consistency constraints.
\newblock In \emph{Proceedings of the 21th ACM SIGKDD International Conference
  on Knowledge Discovery and Data Mining},  635--644.

\bibitem[Little(1993)]{little1993}
Little, R.~J. (1993).
\newblock Statistical analysis of masked data.
\newblock \emph{Journal of Official Statistics} \textbf{9}, 2, 407--426.

\bibitem[Machanavajjhala \emph{et~al.}(2008)Machanavajjhala, Kifer, Abowd,
  Gehrke, and Vilhuber]{machanavajjhala2008}
Machanavajjhala, A., Kifer, D., Abowd, J., Gehrke, J., and Vilhuber, L. (2008).
\newblock Privacy: Theory meets practice on the map.
\newblock In \emph{2008 IEEE 24th International Conference on Data
  Engineering},  277--286. IEEE.

\bibitem[McClure and Reiter(2012)]{mcclure2012}
McClure, D. and Reiter, J.~P. (2012).
\newblock Differential privacy and statistical disclosure risk measures: An
  investigation with binary synthetic data.
\newblock \emph{Trans. Data Priv.} \textbf{5}, 3, 535--552.

\bibitem[McKinney \emph{et~al.}(2020)McKinney, Green, Vilhuber, and
  Abowd]{mckinney2020}
McKinney, K.~L., Green, A.~S., Vilhuber, L., and Abowd, J.~M. (2020).
\newblock Total error and variability measures for the quarterly workforce
  indicators and {LEHD} origin-destination employment statistics in
  {O}n{T}he{M}ap.
\newblock \emph{Journal of Survey Statistics and Methodology}  (online first).

\bibitem[Mervis(2019)]{Mervis2019}
Mervis, J. (2019).
\newblock Researchers object to census privacy measure.
\newblock \emph{Science} \textbf{363}, 6423, 114.

\bibitem[Messing \emph{et~al.}(2020)Messing, DeGregorio, Hillenbrand, King,
  Mahanti, Mukerjee, Nayak, Persily, State, and Wilkins]{Facebook2020}
Messing, S., DeGregorio, C., Hillenbrand, B., King, G., Mahanti, S., Mukerjee,
  Z., Nayak, C., Persily, N., State, B., and Wilkins, A. (2020).
\newblock {Facebook Privacy-Protected Full URLs Data Set}.
\newblock https://doi.org/10.7910/DVN/TDOAPG.

\bibitem[M{\"u}ller and M{\"o}ller(2019)]{muller2019}
M{\"u}ller, D. and M{\"o}ller, J. (2019).
\newblock Giving the international scientific community access to {G}erman
  labor market data: A success story.
\newblock In N.~Crato and P.~Paruolo, eds., \emph{Data-Driven Policy Impact
  Evaluation},  101--117. Springer, Cham.

\bibitem[Nissim \emph{et~al.}(2017)Nissim, Bembenek, Wood, Bun, Gaboardi,
  Gasser, O'Brien, Steinke, and Vadhan]{nissim2017}
Nissim, K., Bembenek, A., Wood, A., Bun, M., Gaboardi, M., Gasser, U., O'Brien,
  D.~R., Steinke, T., and Vadhan, S. (2017).
\newblock Bridging the gap between computer science and legal approaches to
  privacy.
\newblock \emph{Harv. JL \& Tech.} \textbf{31}, 687.

\bibitem[Nissim and Wood(2018)]{nissim2018}
Nissim, K. and Wood, A. (2018).
\newblock Is privacy privacy?
\newblock \emph{Philosophical Transactions of the Royal Society A:
  Mathematical, Physical and Engineering Sciences} \textbf{376}, 2128,
  20170358.

\bibitem[Oberski and Kreuter(2020)]{oberski2020}
Oberski, D.~L. and Kreuter, F. (2020).
\newblock Differential privacy and social science: An urgent puzzle.
\newblock \emph{Harvard Data Science Review} \textbf{2}, 1.

\bibitem[Reiter(2019)]{reiter2019}
Reiter, J.~P. (2019).
\newblock Differential privacy and federal data releases.
\newblock \emph{Annual review of statistics and its application} \textbf{6},
  85--101.

\bibitem[Ruggles \emph{et~al.}(2019)Ruggles, Fitch, Magnuson, and
  Schroeder]{ruggles2019}
Ruggles, S., Fitch, C., Magnuson, D., and Schroeder, J. (2019).
\newblock Differential privacy and census data: Implications for social and
  economic research.
\newblock In \emph{AEA Papers and Proceedings}, vol. 109,  403--408.

\bibitem[Santos-Lozada \emph{et~al.}(2020)Santos-Lozada, Howard, and
  Verdery]{santos2020}
Santos-Lozada, A.~R., Howard, J.~T., and Verdery, A.~M. (2020).
\newblock How differential privacy will affect our understanding of health
  disparities in the {U}nited {S}tates.
\newblock \emph{Proceedings of the National Academy of Sciences} \textbf{117},
  24, 13405--13412.

\bibitem[S{\"a}rndal \emph{et~al.}(2003)S{\"a}rndal, Swensson, and
  Wretman]{sarndal2003}
S{\"a}rndal, C.-E., Swensson, B., and Wretman, J. (2003).
\newblock \emph{Model assisted survey sampling}.
\newblock Springer Science \& Business Media.

\bibitem[Soria-Comas and Drechsler(2013)]{Soria2013}
Soria-Comas, J. and Drechsler, J. (2013).
\newblock Evaluating the potential of differential privacy mechanisms for
  census data.
\newblock In \emph{UNECE work session on statistical data confidentiality}.

\bibitem[Taddicken(2014)]{taddicken2014}
Taddicken, M. (2014).
\newblock The ‘privacy paradox’ in the social web: The impact of privacy
  concerns, individual characteristics, and the perceived social relevance on
  different forms of self-disclosure.
\newblock \emph{Journal of Computer-Mediated Communication} \textbf{19}, 2,
  248--273.

\bibitem[Tang \emph{et~al.}(2017)Tang, Korolova, Bai, Wang, and Wang]{tang2017}
Tang, J., Korolova, A., Bai, X., Wang, X., and Wang, X. (2017).
\newblock Privacy loss in {A}pple's implementation of differential privacy on
  {M}ac{OS} 10.12.
\newblock \emph{arXiv preprint arXiv:1709.02753} .

\bibitem[{Uber Security}(2017)]{Uber2017}
{Uber Security} (2017).
\newblock Uber releases open source project for differential privacy.
\newblock
  https://medium.com/uber-security-privacy/differential-privacy-open-source-7892c82c42b6.

\bibitem[{U.S. Census Bureau}(2020)]{Census2020}
{U.S. Census Bureau} (2020).
\newblock Why your answers matter.
\newblock \url{https://2020census.gov/en/census-data.html}.

\bibitem[{U.S. Census Bureau}(2021{a})]{Census2021}
{U.S. Census Bureau} (2021{a}).
\newblock About the 2020 {C}ensus.
\newblock
  \url{https://www.census.gov/programs-surveys/decennial-census/decade/2020/about.html}.

\bibitem[{U.S. Census Bureau}(2021{b})]{Census_parms}
{U.S. Census Bureau} (2021{b}).
\newblock Census bureau sets key parameters to protect privacy in 2020 census
  results.
\newblock
  \url{https://content.govdelivery.com/accounts/USCENSUS/bulletins/2e32ea9}.

\bibitem[{U.S. Census Bureau}(2021{c})]{Census2021reconstruction}
{U.S. Census Bureau} (2021{c}).
\newblock The {C}ensus {B}ureau's simulated reconstruction-abetted
  re-identification attack on the 2010 {C}ensus.
\newblock
  \url{https://www.census.gov/data/academy/webinars/2021/disclosure-avoidance-series/simulated-reconstruction-abetted-re-identification-attack-on-the-2010-census.html}.

\bibitem[{U.S. Census Bureau}(2021{d})]{Census_quest}
{U.S. Census Bureau} (2021{d}).
\newblock Questionnaires \& instructions.
\newblock
  \url{https://www.census.gov/programs-surveys/decennial-census/technical-documentation/questionnaires.2020_Census.html}.

\bibitem[Vadhan(2017)]{vadhan2017}
Vadhan, S. (2017).
\newblock The complexity of differential privacy.
\newblock In \emph{Tutorials on the Foundations of Cryptography},  347--450.
  Springer.

\bibitem[Van~Riper \emph{et~al.}(2020)Van~Riper, Kugler, and Ruggles]{van2020}
Van~Riper, D., Kugler, T., and Ruggles, S. (2020).
\newblock Disclosure avoidance in the {C}ensus {B}ureau’s 2010 demonstration
  data product.
\newblock In \emph{International Conference on Privacy in Statistical
  Databases},  353--368. Springer.

\bibitem[Warner(1965)]{warner1965}
Warner, S.~L. (1965).
\newblock Randomized response: A survey technique for eliminating evasive
  answer bias.
\newblock \emph{Journal of the American Statistical Association} \textbf{60},
  309, 63--69.

\bibitem[Wasserman and Zhou(2010)]{wasserman2010}
Wasserman, L. and Zhou, S. (2010).
\newblock A statistical framework for differential privacy.
\newblock \emph{Journal of the American Statistical Association} \textbf{105},
  489, 375--389.

\bibitem[Wezerek and Van~Riper(2020)]{NYT2020}
Wezerek, X. and Van~Riper, D. (2020).
\newblock Changes to the census could make small towns disappear.
\newblock
  \url{https://www.nytimes.com/interactive/2020/02/06/opinion/census-algorithm-privacy.html}.

\bibitem[Winkler \emph{et~al.}(2021)Winkler, Butler, Curtis, and
  Egan-Robertson]{winkler2021}
Winkler, R.~L., Butler, J.~L., Curtis, K.~J., and Egan-Robertson, D. (2021).
\newblock Differential privacy and the accuracy of county-level net migration
  estimates.
\newblock \emph{Population Research and Policy Review}  1--19.

\bibitem[Zhu \emph{et~al.}(2021)Zhu, Van~Hentenryck, and Fioretto]{Zhu_2021}
Zhu, K., Van~Hentenryck, P., and Fioretto, F. (2021).
\newblock Bias and variance of post-processing in differential privacy.
\newblock \emph{Proceedings of the AAAI Conference on Artificial Intelligence}
  \textbf{35}, 12, 11177--11184.

\end{thebibliography}

\end{document}